\documentclass[12pt]{article}
\RequirePackage{amsthm,amsmath,amsfonts,amssymb}
\RequirePackage[authoryear]{natbib}
\RequirePackage[colorlinks,citecolor=blue,urlcolor=blue]{hyperref}
\RequirePackage{graphicx}
\usepackage{algorithmic} 
\usepackage[inline]{enumitem}
\usepackage[inline]{enumitem}
\usepackage{booktabs,multirow,threeparttable,graphicx,amsmath,amssymb}
\theoremstyle{plain}

\newtheorem{theorem}{Theorem}[section]

\theoremstyle{definition}

\newtheorem*{example}{Example}

\newtheorem{proposition}{Proposition}
\theoremstyle{remark}
\newtheorem{remark}[theorem]{Remark}

\begin{document}

\renewcommand{\baselinestretch}{1}

\fontsize{12}{14pt plus.8pt minus .6pt}\selectfont \vspace{0.8pc}
\centerline{\large\bf Adaptive Redescending Regression with $q$-Polynomial Losses}
\vspace{2pt} 
\vspace{.4cm} 

 \centerline{Maoxuan Miao$^1$, Shurong Zheng$^1$, You-Gan Wang$^{2*}$ and Jiming Jiang$^3$} 
 
 \centerline{\it $^1$Northeast Normal University}
\centerline{\it $^2$Guangdong University of Finance and Economics and} 
 \centerline{\it $^3$University of California, Davis}
 \normalsize $^{*}$Correspondence: ygwanguq2012@gmail.com\\
\begin{quotation}
\noindent

\begin{abstract}
We introduce a family of smooth $q$-polynomial losses for robust joint
regression--scale estimation. The family contains the Cauchy loss at $q=1$
and converges pointwise to the squared loss as $q$ increases. For every
finite $q$, its regression score is bounded and redescending, while its
scale score is bounded and permits explicit analysis of scale existence.
We formulate simultaneous estimation through a surrogate likelihood with a
common scale calibration $\kappa=1/2$ and select $q$ by maximizing the
normalized surrogate likelihood over a finite candidate set. We investigate
existence, large-sample behaviour, and the robustness implications of the
loss family and discuss how optional modification of the scale score, for
example by taking $\kappa=q/2$, can increase the threshold against scale
explosion. This breakdown-oriented adjustment is distinct from the baseline
estimator and requires its own joint breakdown analysis. An MM-type algorithm
provides a practical implementation. Numerical investigations assess the
resulting robustness--efficiency trade-off under different tail regimes.
\end{abstract}

\vspace{9pt}
\noindent {\it Key words and phrases:}
Asymptotic efficiency, breakdown points, consistency, $M$-estimation, optimal estimating functions.
\par
\end{quotation}\par

\def\thefigure{\arabic{figure}}
\def\thetable{\arabic{table}}

\renewcommand{\theequation}{\thesection.\arabic{equation}}

\fontsize{12}{14pt plus.8pt minus .6pt}\selectfont



\section{Introduction}

The least squares estimator (LSE) is the classical benchmark for estimating the
regression coefficient $\beta$. When the regression errors $\varepsilon_i$ are
light-tailed and approximately Gaussian, the LSE is statistically efficient.
Under heavy-tailed error distributions or in the presence of outliers, however,
the LSE can be highly unstable and may deviate substantially from the target
regression coefficient. This limitation has motivated a large literature on
robust regression methods that retain reliable performance under departures
from classical distributional assumptions.

A broad and well-studied class of robust procedures is provided by
M-estimators, which replace the squared loss by a loss function with slower tail
growth. Classical examples include the least absolute deviation (LAD) estimator,
whose loss grows linearly in the residuals \citep{pollard1991asymptotics}. The
global robustness of an M-estimator is often assessed through its finite-sample
breakdown point, defined as the smallest fraction of contaminated observations
that can force the estimator to take arbitrarily large or otherwise aberrant
values \citep{hampel1971general}. In fixed-design regression, the breakdown
behavior of an M-estimator depends critically on the tail growth of its loss
function. Losses with slower tail growth generally provide greater resistance to
extreme outliers. For example, Cauchy-type estimators can attain the highest
asymptotic breakdown point, whereas estimators based on linear-growth losses,
such as LAD, Huber, and $\ln\!\cosh$, have substantially weaker
breakdown-point properties. This observation highlights the strong robustness of
Cauchy-type losses under extremely heavy-tailed errors.

These insights have led to extensive work on robust regression methods with
high breakdown points under contamination in both the response and covariate
spaces. Broadly, two main strategies have emerged. The first enhances
robustness through the scale estimator. A classical example is the class of
S-estimators, which achieve high breakdown points by minimizing a robust scale
functional of the residuals \citep{rousseeuw1984robust}. Because their
robustification parameter is fixed, however, S-estimators typically sacrifice
efficiency under Gaussian errors. The MM estimator of \citet{yohai1987high}
addresses this issue through a two-stage construction: an initial
high-breakdown estimator is followed by an efficiency-improving M-estimation
step. Related developments include the $\tau$-estimator
\citep{yohai1988high} and modified one-step M-estimators
\citep{coakley1993bounded}. Although these methods improve efficiency relative
to S-estimators, the coupling of tuning parameters can limit their ability to
maintain strong robustness under heavy-tailed errors while achieving high
efficiency near the Gaussian model.

A second line of work adopts simultaneous regression--scale estimation, which
arises naturally from likelihood-based formulations. A conceptually important
development in this direction is the robust estimation framework of
\citet{Catoni2012}, which emphasizes explicit control of tail behavior through
bounded influence and exponential-type inequalities. Catoni's work provides a
principled mechanism for balancing robustness and efficiency and has influenced
a wide range of modern robust procedures. In regression settings,
likelihood-based estimators derived from heavy-tailed distributions, such as
the Cauchy or low-degree-of-freedom $t$ distributions, can achieve high
breakdown points through suitable likelihood tuning. However, such estimators
often have low efficiency under near-Gaussian errors, and their robustness may
deteriorate as the model is tuned toward greater efficiency.

More generally, many robust M-estimators employ bounded or slowly increasing
loss functions with tuning parameters chosen to balance efficiency and
robustness. Careful tuning can yield high efficiency under Gaussian errors while
preserving a positive breakdown point. Nevertheless, moving the estimator closer
to least-squares behavior typically weakens protection against heavy-tailed
errors and gross contamination. It remains challenging to construct a
regression--scale estimator that adapts to unknown tail behavior, achieves high
efficiency under Gaussian or near-Gaussian errors, and retains strong global
robustness, including a high breakdown point, under heavy-tailed errors.

Suppose that we observe data $\{(x_i,y_i)\}_{i=1}^n$ generated from the linear
model
\begin{equation}
    y_i = x_i^{\top}\beta + \varepsilon_i, \qquad i=1,\ldots,n,
    \label{line_mode}
\end{equation}
where $y_i\in\mathbb{R}$, $x_i\in\mathbb{R}^p$, and the errors
$\varepsilon_i$ are independent and identically distributed and independent of
the covariates. This paper addresses the following question:
\emph{can one construct an adaptive regression--scale estimator that achieves
a data-adaptive transition between redescending robustness and the
least-squares limit, with tractable joint regression--scale inference?}

Our starting point is a new family of $q$-polynomial losses. The index $q$
governs a transparent transition from the Cauchy loss to the squared loss:
for finite $q$ the score redescends, whereas increasing $q$ makes the loss
more nearly quadratic on a growing range of residual values. Rather than
impose a fixed robustness level, we select $q$ over a finite candidate set
by maximizing a normalized surrogate likelihood. The surrogate family is an
estimation device and need not coincide with the true error distribution.
Likelihood-based estimation under model misspecification is classical
\citep{white:1982}. More closely related to robust regression,
\citet{fu:wang:cai:2020} proposed a working-likelihood approach to
calibrating robustness tuning, building on data-dependent Huber tuning
\citep{wang2007robust}; see also \citet{wu2023working}.
Here we use the term \emph{surrogate likelihood} to emphasize that the
polynomial-induced density is an estimation device, not an assumed true
error distribution.

For the main development, we fix the coefficient of the log-scale term at
$\kappa=1/2$. This yields a common likelihood-based formulation for every
finite $q$ and reduces the tuning problem to the choice of the loss index.
Motivated by the joint regression--scale formulation of
\citet{he2000breakdown} and the tuned Cauchy approach of \citet{mizera2002breakdown}, we
also examine the role of the scale equation in existence and robustness.
In particular, a separate breakdown-oriented calibration, such as
$\kappa=q/2$, increases the fixed-regression threshold against scale
explosion, although a joint breakdown guarantee requires a separate proof.

The proposed construction is also compatible with high-dimensional
regularization. In particular, the empirical regression--scale objective can be
augmented by a sparsity-inducing penalty, such as the lasso, SCAD, MCP, or
related penalties, to form penalized robust estimators for large-$p$ model
selection problems. From this perspective, the $q$-polynomial loss provides an
adaptive robust data-fitting component, while the penalty controls model
complexity. The present paper develops the unpenalized regression--scale theory,
which provides a basic building block for such extensions: loss properties,
scale existence, asymptotic inference under suitable regularity conditions,
and data-driven selection of $q$. High-dimensional penalized versions can
therefore be built on the proposed loss-and-scale framework, although their full
analysis would require additional tools involving restricted eigenvalue
conditions, penalty bias, and post-selection behavior.

The main contributions are the analytical characterization of the new
$q$-polynomial loss and its regression and scale scores, a unified
surrogate-likelihood formulation for joint regression--scale estimation,
and theory for data-driven selection of $q$ over a finite candidate class.
We distinguish these statistical properties from the convergence properties
of the nonconvex numerical algorithm. The optional likelihood adjustment
of the scale score is discussed separately as a means of strengthening
breakdown protection, rather than as a second parameter estimated from the
data.

\section{Adaptive Regression--Scale Polynomial Estimator}
We first characterize the new loss function and its associated estimating
functions. We then derive joint regression--scale estimation from a
normalized surrogate likelihood, keeping the log-scale coefficient fixed
at $\kappa=1/2$ in the principal methodology.

\medskip

To operationalize this idea, we consider a family of polynomial-based losses
indexed by $q\ge1$,
$\ell_q(u)
=
\log\!\Biggl(\sum_{t=0}^{q}\frac{u^{2t}}{t!}\Biggr).$
The index $q$ controls tail growth: small $q$ produces a slowly increasing loss
with strong protection against extreme outliers, whereas large $q$ yields a
criterion that is increasingly quadratic near the origin and therefore more
efficient under Gaussian or near-Gaussian errors.

\begin{remark}
When $q=1$, $\ell_1(u)=\log(1+u^2)$ coincides with the Cauchy loss, which is
well suited to extremely heavy-tailed errors. As $q\to\infty$, the truncated
series converges to $\exp(u^2)$ and hence $\ell_q(u)\to u^2$, recovering the
least-squares loss. Thus, the family $\{\ell_q\}_{q\ge1}$ provides a smooth
bridge between highly robust Cauchy-type estimation and efficient least-squares
estimation.
\end{remark}

Let $\psi_q(u)=\tfrac12\ell_q'(u)$ denote the associated score function, which
admits the explicit form
$\psi_q(u)
=
\frac{
u \sum_{t=0}^{q-1}\frac{u^{2t}}{t!}
}{
\sum_{t=0}^{q}\frac{u^{2t}}{t!}
}.$

\begin{proposition}[Analytical properties of the polynomial losses]
\label{prop:polynomial_loss_properties}
For every positive integer $q$, the loss $\ell_q$ is even,
nonnegative, continuously differentiable and strictly increasing in $|u|$.
Writing $\psi_q(u)=\ell_q'(u)/2$ and $\chi_q(u)=u\psi_q(u)$,
we have, as $|u|\to\infty$,
\[
\ell_q(u)=2q\log|u|-\log(q!)+o(1),\qquad
\psi_q(u)=\frac{q}{u}+o(|u|^{-1}),\qquad
\chi_q(u)\longrightarrow q.
\]
The regression score $\psi_q$ is bounded and redescending,
and $\chi_q$ is strictly increasing in $|u|>0$ from zero to $q$.
For each fixed $u$, $\ell_q(u)\to u^2$ as $q\to\infty$.
Moreover, $\ell_q$ is nonconvex for every finite $q$.
\end{proposition}

\begin{proof}
Let $P_q(z)=\sum_{t=0}^q z^t/t!$. Direct differentiation yields
$\psi_q(u)=uP_{q-1}(u^2)/P_q(u^2)$.
The tail expansions follow from $P_q(z)\sim z^q/q!$ and
$P_{q-1}(z)\sim z^{q-1}/(q-1)!$.
To establish monotonicity of $\chi_q$, assign weights
$w_t(z)=(z^t/t!)/P_q(z)$ to $t\in\{0,\ldots,q\}$.
Then $\chi_q(\sqrt z)=\sum_t t w_t(z)$ and its derivative
with respect to $\log z$ equals the strictly positive variance
of the corresponding discrete random variable for $z>0$.
The limit $\ell_q(u)\to u^2$ follows from the exponential series.
Finally, $\psi_q(u)$ is positive for $u>0$ but tends to zero as
$u\to\infty$, which excludes global convexity of $\ell_q$.
\end{proof}

\medskip

The influence function characterizes the local robustness of an estimator by quantifying the infinitesimal effect of a single observation \(z=(\tilde x,\tilde y)\) on the estimator; see \citet{hampel1974influence,daouia2006robust,mathieu2022concentration}. For each finite $q$, the regression score $\psi_q(u)$ is bounded and
redescends to zero as $|u|\to\infty$. This provides bounded response
influence for bounded designs under the usual nonsingularity conditions.
Because the regression estimating equation contains the factor $x$, however,
bounded influence against arbitrary leverage contamination does not follow
without additional design restrictions or leverage weights. The working
likelihood analysis of \citet{he2000breakdown} is especially relevant to
this distinction.

\begin{figure}[tbp]
    \centering
    \includegraphics[width=1\linewidth]{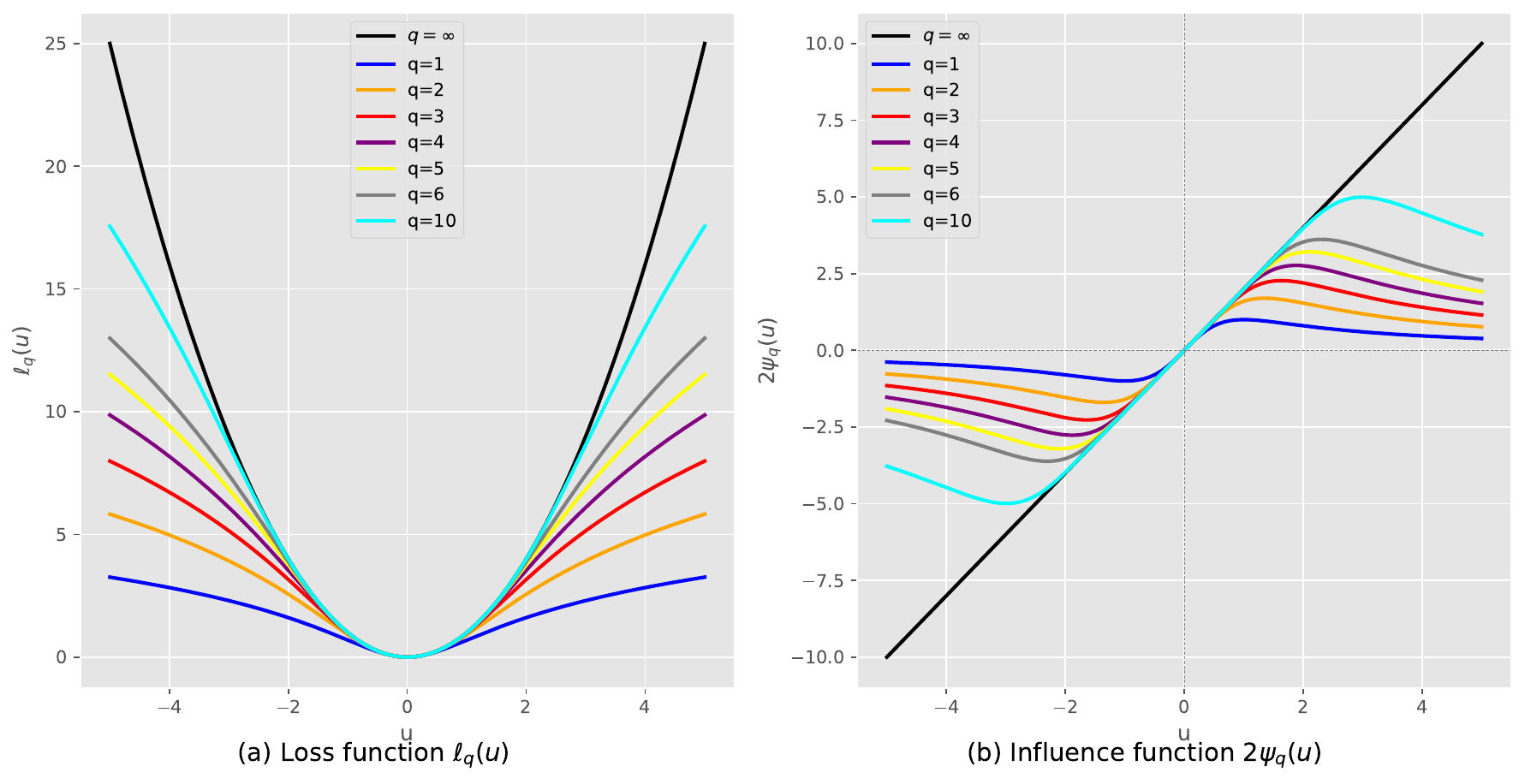}
    \caption{$\ell_q(u)$ and $2\psi_q(u)$ as functions of $u$.}
    \label{loss_influenc}
\end{figure}

Figure~\ref{loss_influenc} displays $\ell_q(u)$ and the corresponding $2\psi_q(u)$
for several values of $q$. As $q$ increases, $\ell_q$ becomes nearly quadratic
near the origin and the score $2\psi_q$ is close to linear over a wider range,
indicating improved efficiency under Gaussian errors. In contrast, small $q$
induces pronounced tail damping in both the loss and the score, yielding strong
downweighting of extreme residuals and enhanced robustness to outlying
responses.

\medskip

To jointly estimate the regression coefficient $\beta$ and scale parameter
$\sigma$, we start from the surrogate density
$f_q(u)=\{c(q)P_q(u^2)\}^{-1}$, where
$P_q(z)=\sum_{t=0}^q z^t/t!$ and
$c(q)=\int_{\mathbb R}P_q(u^2)^{-1}\,du<\infty$.
This density is proper for every finite integer $q\geq1$ and need not be
the true error density. Its location--scale likelihood motivates
\begin{equation}
    \hat{\mathcal{L}}_{n}(\beta,\sigma \mid \kappa,q)
    =
    \frac{1}{n}\sum_{i=1}^n
    \left\{
        \frac{1}{2}\ell_q\!\left(
        \frac{y_i-x_i^{\top}\beta}{\sigma}\right)
        + \kappa \log \sigma
    \right\},
    \label{empiri_risk}
\end{equation}
where the default is $\kappa=1/2$. At that value, twice the criterion,
plus $\log c(q)$, equals the average negative log surrogate likelihood.
The resulting baseline estimates of regression and scale are therefore
obtained without estimating a separate tuning coefficient.

The idea of modifying likelihood scores has several precedents:
\citet{firth1993bias} adjusted scores to reduce first-order bias,
whereas \citet{he2000breakdown} studied robust weighted
working-likelihood estimation. Of particular relevance,
\citet{mizera2002breakdown} adjusted the coefficient of the
log-scale term of the Cauchy criterion to investigate breakdown.
Their tuned criterion is recovered from \eqref{empiri_risk} when $q=1$,
up to a positive multiplicative factor. In our setting $\kappa$
shifts the scale estimating equation but does not alter the functional
form of the regression score. We retain $\kappa=1/2$ for the main
estimator and discuss $\kappa=q/2$ only as an optional
breakdown-oriented adjustment.

For $q$ and a fixed calibration $\kappa$ (with $\kappa=1/2$ by default),
the regression--scale polynomial estimator $(\hat{\beta}_{n,q,\kappa},
\hat{\sigma}_{n,q,\kappa})$ is defined, whenever a global
minimizer exists, by
\begin{equation}
    (\hat{\beta}_{n,q,\kappa},\hat{\sigma}_{n,q,\kappa})
    \in
    \arg\min_{(\beta,\sigma)\in\mathbb{R}^p\times(0,\infty)}
    \hat{\mathcal{L}}_{n}(\beta,\sigma \mid \kappa,q).
\end{equation}
Any interior minimizer necessarily satisfies the system
of estimating equations
\begin{align}
    \frac{1}{n}\sum_{i=1}^n
    \psi_q\!\left(
    \frac{y_i-x_i^{\top}\beta}{\sigma}\right)x_i
    &= 0,
    \label{sigma_solu2}
    \\
    \frac{1}{n}\sum_{i=1}^n
    \chi_q\!\left(
    \frac{y_i-x_i^{\top}\beta}{\sigma}\right)
    &= \kappa,
    \label{sigma_solu}
\end{align}
where
$
\chi_q(u)
=
u\,\psi_q(u)
=
\frac{
    u^2 \sum_{t=0}^{q-1} u^{2t}/t!
}{
    \sum_{t=0}^{q} u^{2t}/t!
}.
$

\medskip

\begin{proposition}[Conditional scale existence and uniqueness]
\label{prop:scale_solution}
For fixed $\beta\in\mathbb R^p$, let
$m(\beta)=\#\{i:y_i-x_i^\top\beta\ne0\}$.
The scale equation \eqref{sigma_solu} has a unique positive solution
if and only if $0<\kappa<qm(\beta)/n$. In this case the
solution uniquely minimizes \eqref{empiri_risk} over $\sigma>0$
for this fixed $\beta$. In particular, at $\kappa=1/2$ a unique
positive scale solution exists if and only if
$m(\beta)>n/(2q)$.
\end{proposition}

\begin{proof}
Put $\eta=\log\sigma$. The first derivative of the objective
with respect to $\eta$ is
$\kappa-n^{-1}\sum_i\chi_q\{(y_i-x_i^\top\beta)e^{-\eta}\}$.
By Proposition~\ref{prop:polynomial_loss_properties}, this derivative
is strictly increasing in $\eta$ if $m(\beta)>0$, and its limits
at $-\infty$ and $+\infty$ are
$\kappa-qm(\beta)/n$ and $\kappa$, respectively.
It therefore has a unique zero exactly under the asserted
inequality. Strict convexity in $\eta$ gives uniqueness and
conditional global minimality. If $m(\beta)=0$, the scale
score is identically zero and no solution exists for $\kappa>0$.
\end{proof}

Conditional scale uniqueness does not imply uniqueness of the
joint regression--scale minimizer, since the objective is generally
nonconvex in $\beta$.

Let $F$ denote the joint distribution of $(x,y)$. The population target
$(\beta_{q,\kappa}^*,\sigma_{q,\kappa}^*)$ is defined as
$
(\beta_{q,\kappa}^*,\sigma_{q,\kappa}^*)
=
\arg\min_{(\beta,\sigma)}
\mathbb{E}_F
\left\{
    \frac{1}{2}\ell_q\!\left(\frac{y-x^{\top}\beta}{\sigma}\right)
    + \kappa \log \sigma
\right\},
$
and satisfies
\begin{equation}
\begin{aligned}
\mathbb{E}_F\!\left[
\psi_q\!\left(\frac{y-x^{\top}\beta}{\sigma}\right)x
\right] &= 0, \\
\mathbb{E}_F\!\left[
\chi_q\!\left(\frac{y-x^{\top}\beta}{\sigma}\right)
\right] &= \kappa.
\end{aligned}
\label{Scal_beta}
\end{equation}
The dependence of $(\beta_{q,\kappa}^*,\sigma_{q,\kappa}^*)$ on $q$ reflects the fact that the
effective target of estimation adapts to the tail behavior implied by the loss.
In subsequent sections, we show that selecting $q$ in a data-driven manner yields
an estimator that adapts automatically to unknown error distributions while
retaining desirable robustness and efficiency properties.

Section~2 defined the regression--scale polynomial estimator through the
nonconvex objective~\eqref{empiri_risk} and the associated estimating
equations~\eqref{sigma_solu2}--\eqref{sigma_solu}. Before turning to asymptotic
properties and robustness, we first establish that the estimator is
\emph{well-defined}, namely that the empirical objective admits at least one
(global) minimizer over $\Theta=\mathbb{R}^p\times(0,\infty)$. Since the scale
equation~\eqref{sigma_solu} depends on the likelihood-adjustment parameter
$\kappa$, an inappropriate choice of $\kappa$ may destroy existence. A detailed discussion of the sufficient conditions on $\kappa$ under which
\eqref{empiri_risk} admits a global minimizer is provided in Section S2 of the
Supplementary Material.

\medskip


\section{Asymptotic Properties}\label{sec:asy}
This section establishes consistency and asymptotic normality of the
regression--scale polynomial estimator. The results follow from standard
$M$-estimation theory for joint regression--scale estimation; see, for example,
\citet{maronna1981asymptotic}. We restrict attention to a class of distributions $F$ under which the population
estimating equations~\eqref{Scal_beta} admit a unique solution
$(\beta_{q,\kappa}^*,\sigma_{q,\kappa}^*)$; see \citet{maronna1981asymptotic} and Section~S3 of the Supplementary Material for further details. The theoretical results in the Supplementary Material, Theorems~S3.1 and~S3.2 show that for a fixed $q$ the
regression--scale polynomial estimator is strongly consistent and asymptotically
normal for the population target $(\beta_{q,\kappa}^*,\sigma_{q,\kappa}^*)$. If $q$ is poorly
matched to the error distribution, the target $(\beta_{q,\kappa}^*,\sigma_{q,\kappa}^*)$ may
deviate from the true parameter $(\beta,\sigma)$, increasing the approximation
error $\|(\beta,\sigma)-(\beta_{q,\kappa}^*,\sigma_{q,\kappa}^*)\|_2$. This motivates data-driven
selection of $q$ to mitigate the approximation error; see
Section~\ref{sec6}.

\section{Breakdown implications of scale calibration}\label{BP}
For the main estimator, the scale-score equation is
$n^{-1}\sum_i\chi_q(r_i/\sigma)=1/2$. For each fixed
regression coefficient $\beta$, the scale score $\chi_q$ is bounded,
with $\chi_q(u)\to q$ as $|u|\to\infty$. These properties
make it possible to evaluate the threshold for scale explosion under
vertical contamination conditional on $\beta$.

If a fraction $\varepsilon$ of the residuals diverges while the
remaining residuals stay bounded, then a finite limiting scale cannot
solve the baseline equation when $\varepsilon q>1/2$. Thus the
conditional scale-explosion threshold is governed by $1/(2q)$.
This conditional calculation should not be confused with the
finite-sample breakdown point of the jointly optimized regression--scale
estimator. Joint breakdown also depends on the number of exact fits,
the geometry of the design, and the possibility of scale implosion;
see \citet{mizera2002breakdown} and \citet{he2000breakdown}.
The latter authors' leverage-robustness conclusions require
additional weights and cannot be transferred unchanged to the
unweighted estimator considered here.

\section{A Breakdown-Oriented Extension:
Scale Calibration with $\kappa=q/2$}
\label{sec:breakdown_extension}

Throughout the main development, we fix $\kappa=1/2$
and select the robustness parameter $q$ by maximizing
the surrogate likelihood. This formulation provides
a common likelihood-based criterion and allows the
loss to approach the least-squares loss as $q$ increases.
However, the corresponding robustness against scale
explosion decreases with $q$. Motivated by the
tuned-likelihood approach of \citet{mizera2002breakdown},
we investigate an alternative calibration,
$\kappa=q/2$, which provides stronger protection
against extreme residuals.

\subsection{Adjusted surrogate likelihood}

Under $\kappa=q/2$, the scale estimating equation
becomes
\begin{equation}
    \frac{1}{n}\sum_{i=1}^n
    \chi_q\!\left(
    \frac{y_i-x_i^\top\beta}{\sigma}
    \right)
    =\frac q2.
    \label{eq:adjusted_scale_equation}
\end{equation}
For fixed $\beta$, this adjustment increases the
critical fraction of extreme residuals required
to force scale explosion from $1/(2q)$ to $1/2$.

The adjusted objective is proportional, up to an
additive constant, to the negative log-likelihood
associated with the surrogate density
\begin{equation}
    f_q^{\mathrm{adj}}(u)
    =
    \frac{1}{Z_q}
    \left\{
    \sum_{t=0}^{q}\frac{u^{2t}}{t!}
    \right\}^{-1/q},
    \label{eq:adjusted_density}
\end{equation}
where $Z_q$ is the normalizing constant.
For every finite integer $q\geq1$, this density
has tails of order $|u|^{-2}$. In particular,
$q=1$ recovers the standard Cauchy density.

Although $\ell_q(u)\to u^2$ for each fixed $u$,
the adjusted density does not converge to a
Gaussian distribution on the original scale.
Instead, its dispersion increases with $q$,
motivating a rescaled representation.

\subsection{Limiting distribution under rescaling}

Let $U_q$ have density $f_q^{\mathrm{adj}}$,
and define $V_q=U_q/\sqrt q$. Its density is
\begin{equation}
    g_q(v)
    =
    \frac{1}{C_q}
    \left\{
    \sum_{t=0}^{q}\frac{(qv^2)^t}{t!}
    \right\}^{-1/q},
    \label{eq:rescaled_adjusted_density}
\end{equation}
where $C_q=Z_q/\sqrt q$.

\begin{theorem}[Limiting adjusted surrogate density]
\label{thm:adjusted_limit}
As $q\to\infty$, the density $g_q$ converges
in $L^1(\mathbb R)$ to
\begin{equation}
    g_\infty(v)
    =
    \frac{1}{C_\infty}
    \begin{cases}
        \exp(-v^2), & |v|\leq1,\\
        e^{-1}|v|^{-2}, & |v|>1,
    \end{cases}
    \label{eq:adjusted_limit_density}
\end{equation}
where
$
C_\infty
=
\sqrt{\pi}\operatorname{erf}(1)+2/e.
$
Consequently, $V_q$ converges in distribution
to a random variable with a Gaussian-shaped
central density and Cauchy-order tails.
\end{theorem}

\begin{proof}
Let $P_q(z)=\sum_{t=0}^q z^t/t!$.
The asymptotic behaviour of the truncated
exponential series gives
\[
    \lim_{q\to\infty}
    \frac{1}{q}\log P_q(qv^2)
    =
    \begin{cases}
        v^2, & |v|\leq1,\\
        1+\log(v^2), & |v|>1.
    \end{cases}
\]
Hence the unnormalized densities converge
pointwise to the kernel in
\eqref{eq:adjusted_limit_density}.

Furthermore, these kernels are bounded by
one on $|v|\leq1$ and by $|v|^{-2}$
on $|v|>1$, using
$P_q(qv^2)\geq(qv^2)^q/q!$.
The dominating function is integrable.
Dominated convergence therefore establishes
convergence of the normalizing constants and
the normalized densities in $L^1$.
\end{proof}

The corresponding limiting loss, apart from
an additive constant, is
\begin{equation}
    \rho_\infty(v)
    =
    \begin{cases}
        v^2, & |v|\leq1,\\
        1+2\log|v|, & |v|>1.
    \end{cases}
    \label{eq:adjusted_limiting_loss}
\end{equation}
Its scale-score function is
$\chi_\infty(v)=\min\{v^2,1\}$.

Writing $\tau=\sqrt q\,\sigma$, the rescaled
residual is $v=(y-x^\top\beta)/\tau$.
The limiting loss is therefore quadratic
for residual magnitudes not exceeding $\tau$
and logarithmic beyond this threshold.
This provides a well-defined robustification
threshold as $q$ increases.

\subsection{Implications for scale breakdown}

A notable feature of the adjusted family is
that the two endpoint cases, $q=1$ and
$q\to\infty$, exhibit the same critical
contamination fraction for scale explosion
when the regression coefficient is fixed.

\begin{proposition}[Conditional scale-explosion threshold]
\label{prop:adjusted_scale_breakdown}
Fix $\beta$ and suppose that $m$ residual
magnitudes diverge while the remaining
$n-m$ residuals stay fixed and bounded.
Consider either the adjusted scale equation
at a finite integer $q\geq1$ or the
limiting scale equation induced by
\eqref{eq:adjusted_limiting_loss}.

If $m<n/2$, positive scale solutions remain
bounded above under this contamination.
If $m>n/2$, every positive scale solution
diverges. Thus the critical contamination
fraction for conditional scale explosion
is $1/2$, with the exact half-sample
boundary requiring separate consideration.
\end{proposition}

\begin{proof}
For finite $q$, divide
\eqref{eq:adjusted_scale_equation} by $q$
and define $h_q(v)=\chi_q(\sqrt q\,v)/q$,
using $\tau=\sqrt q\,\sigma$.
For the limiting case, define
$h_\infty(v)=\min\{v^2,1\}$.

In either case, the scale equation has the
form
\[
    \frac1n\sum_{i=1}^n
    h(|r_i|/\tau)=\frac12,
\]
where $h$ is continuous, nondecreasing,
bounded by one, satisfies $h(0)=0$,
and approaches one at infinity.

When $m<n/2$, choose a sufficiently large
fixed $\tau$ so that the contributions from
the uncontaminated residuals are smaller
than $1/2-m/n$. The scale equation cannot
hold at any larger value of $\tau$, so
scale explosion is impossible.

When $m>n/2$, the contributions from the
diverging residuals alone eventually
exceed $1/2$ at every bounded positive
value of $\tau$. Therefore any positive
scale solution must diverge.
\end{proof}

The result concerns scale explosion conditional
on a fixed regression coefficient. It does not
establish a $50\%$ breakdown point for the
joint regression--scale estimator, which also
depends on the design configuration, scale
implosion, and global minimization.

The calibration $\kappa=q/2$ is therefore
an interesting robustness-oriented alternative
to the baseline $\kappa=1/2$. We retain the
latter for the principal estimation procedure,
while a complete characterization of joint
breakdown under the adjusted family is left
for future research.

\section{Maximum Surrogate-Likelihood Selection of $q$}
\label{sec6}
We select the loss index $q$ over the finite candidate set
$\mathcal M=\{1,\ldots,q_{\max}\}$ by maximizing the surrogate
likelihood. We do not estimate $\kappa$ from this likelihood:
$\kappa=1/2$ is fixed throughout both loss selection and final fitting.
This distinction is important because $\kappa$ is a scale-score
calibration, not an additional parameter of the normalized surrogate
density defined below.

For each $q\in\mathcal M$, the standardized surrogate density is
\begin{equation}\label{eq:work_density}
    f_q(u)=\frac{1}{c(q)}
      \left\{\sum_{t=0}^{q}\frac{u^{2t}}{t!}\right\}^{-1},
    \qquad
    c(q)=\int_{\mathbb R}
      \left\{\sum_{t=0}^{q}\frac{u^{2t}}{t!}\right\}^{-1}du.
\end{equation}
The negative log-likelihood contribution at residual
$r=(y-x^\top\beta)/\sigma$ is
\begin{equation}\label{eq:negloglik}
    -\log\{\sigma^{-1}f_q(r)\}
    =\ell_q(r)+\log c(q)+\log\sigma.
\end{equation}
Let $(\hat\beta_{n,q,1/2},\hat\sigma_{n,q,1/2})$
denote a fitted global minimizer of \eqref{empiri_risk} with
$\kappa=1/2$, whenever one exists. We use the profiled criterion
\begin{equation}\label{eq:Hn_def}
 H_n(q)=\frac{1}{n}\sum_{i=1}^n
 \ell_q\!\left(\frac{y_i-x_i^\top\hat\beta_{n,q,1/2}}
 {\hat\sigma_{n,q,1/2}}\right)
 +\log c(q)+\log\hat\sigma_{n,q,1/2},
\end{equation}
and choose
\begin{equation}\label{eq:qhat_def}
    \hat q\in\arg\min_{q\in\mathcal M}H_n(q).
\end{equation}
The $q$-dependent normalizing constant $c(q)$ is essential
when comparing candidate losses. Since every candidate has the
same $p+1$ fitted regression and scale parameters, an information
criterion with a common parameter-count penalty yields the same
ranking. We therefore refer simply to maximum surrogate-likelihood
selection rather than introducing an AIC-type rule.

To interpret selection, let $(\beta_{q,1/2}^*,\sigma_{q,1/2}^*)$
minimize the population surrogate negative log-likelihood and set
\begin{equation}\label{eq:qstar_KL_new}
    s(q)=E\!\left[
      \ell_q\!\left(\frac{Y-X^\top\beta_{q,1/2}^*}
      {\sigma_{q,1/2}^*}\right)\right]
      +\log c(q)+\log\sigma_{q,1/2}^*,
      \qquad q_*\in\arg\min_{q\in\mathcal M}s(q).
\end{equation}
When the true conditional distribution is not in the surrogate
family, $q_*$ is the population KL-projection index, provided the
expected log likelihood is well-defined. A unique minimizer
$q_*$ together with uniform convergence of the profiled
criteria on the finite set yields consistent selection.

\begin{theorem}[Consistency of the surrogate-likelihood selector]
\label{q_conver}
Suppose that $\mathcal M$ is finite, $q_*$ is the unique
minimizer of $s(q)$, and
$H_n(q)\xrightarrow{P}s(q)$ for each $q\in\mathcal M$.
Then $\Pr(\hat q=q_*)\to1$.
\end{theorem}

\begin{proof}
Because $\mathcal M$ is finite and $q_*$ is the unique
minimizer, $\min_{q\ne q_*}\{s(q)-s(q_*)\}>0$.
Convergence in probability of all finitely many criteria
implies that, with probability tending to one, their errors
are simultaneously smaller than half this gap.
On this event $\hat q=q_*$.
\end{proof}

\begin{theorem}[Oracle equivalence after selecting $q$]
\label{thm:adaptive_q}
Assume the conditions of Theorem~\ref{q_conver} and suppose
that for fixed $q_*$ the baseline estimator
$(\hat\beta_{n,q_*,1/2},\hat\sigma_{n,q_*,1/2})$
is consistent for $\theta_{q_*}^*$ and satisfies
\[
\sqrt n\{(\hat\beta_{n,q_*,1/2},\hat\sigma_{n,q_*,1/2})
-\theta_{q_*}^*\}\xrightarrow{d}N_{p+1}(0,V_{q_*}).
\]
Then the plug-in estimator
$(\hat\beta_{n,\hat q,1/2},\hat\sigma_{n,\hat q,1/2})$
has the same consistency and limiting distribution.
\end{theorem}

\begin{proof}
By Theorem~\ref{q_conver},
$\Pr(\hat q=q_*)\to1$. Hence the plug-in
estimator equals its fixed-$q_*$ counterpart with probability
tending to one. The asserted convergence results follow.
\end{proof}

The scale parameter in these statements is a surrogate-likelihood
scale, which need not equal the physical standard deviation under
model misspecification. In the Gaussian limit at fixed
$\kappa=1/2$, the surrogate scale corresponds to
$\sqrt{2}$ times the Gaussian standard deviation. The baseline
choice $\kappa=1/2$ is maintained after selecting $\hat q$;
there is no second-stage refitting at $\kappa(\hat q)$.

\section{Algorithm for Implementation}\label{sec:algo_impl}

Since Eqs.~\eqref{sigma_solu2} and~\eqref{sigma_solu} do not admit closed-form
solutions, we compute \((\hat\beta_{n,q,\kappa},\hat\sigma_{n,q,\kappa})\) by a two-block
iterative scheme in the spirit of \citet{green1984iteratively}. For nonconvex
losses, a direct scale update may fail to produce a descent sequence
\citep{shanno1986numerical,wolke1988iteratively,huber1992robust}. We therefore
combine an IRLS update for the regression coefficient with a local adaptive
majorize--minimization (LAMM) update for the scale parameter
\citep{fan2018lamm,fan2021provable}. The resulting procedure is referred to as
an MM-based block coordinate descent algorithm.

At iteration \(r\), given \(\hat\sigma^{(r)}\), the regression coefficient
\(\hat\beta^{(r+1)}\) is updated by IRLS. Given \(\hat\beta^{(r+1)}\), the scale
parameter \(\hat\sigma^{(r+1)}\) is updated by LAMM with backtracking, so that
the empirical objective is nonincreasing at each iteration. The iterations are
stopped when the norm of the joint estimating equations for \((\beta,\sigma)\)
falls below a prescribed tolerance \(\varepsilon'\). In particular, the algorithm
generates a descent sequence of objective values and targets an
\(\varepsilon'\)-approximate stationary point of the empirical problem.

The complete update formulas, the stopping criterion, and the monotone descent
result are given in Section~S6 of the Supplementary Materials.

\section{Numerical Investigation}
\noindent\emph{Revision note.} The numerical results retained below are
from the previous estimator implementation. While the original
likelihood selector was equivalent to maximum surrogate likelihood
because the model dimensions were equal, several reported results
used a subsequent $q$-dependent scale calibration. Those values are
preserved for comparison and must be recomputed for the revised
fixed-$\kappa=1/2$ estimator before submission.

\subsection{Simulation studies}

In this section, we conduct simulation studies to evaluate the finite-sample
performance of the proposed estimator and to examine maximum surrogate-likelihood
selection of the robustness parameter $q$. Two examples are considered.

\begin{example}
\textbf{Surrogate-likelihood selection of $q$ (original numerical results).}
We assess the original numerical results for selecting the robustness parameter $q$ by the normalized surrogate likelihood.
We generate $1000$ datasets from the linear regression model
\[
y_i = \beta_0 + \beta_1 x_{i1} + \varepsilon_i,\qquad i=1,\ldots,n,
\]
with $n=400$, $\beta_0=1$, $\beta_1=2$, and $x_{i1}\sim N(0,1)$.
The errors $\varepsilon_i$ are generated from Student’s $t(\nu)$ distributions
with $\nu \in \{1,\,1.5,\,1.9,\,2.5,\,4.3,\,25,\,26\}$, as well as from the
standard normal distribution $N(0,1)$.

We compare the adaptive regression--scale polynomial estimator with $q$ selected
by surrogate likelihood (originally labeled ARSPE--AIC) to regression--scale polynomial estimators with fixed
$q\in\{1,\ldots,10\}$.  We set $q_{\max}=10$, $\gamma_u = 1.5$, $\hat{\beta}^{(0)}=(1,1)$, $\hat{\sigma}^{(0)}=0.5$,
$\phi_0=10^{-4}$, $\varepsilon'=10^{-3}$, and use $\kappa=0.5$ for all $q$.

Performance is evaluated using mean squared error (MSE), absolute bias, and the
average selected robustness parameter
$
\bar{q} = \frac{1}{1000}\sum_{b=1}^{1000} \hat q_b,
$
where $\hat q_b$ is the surrogate-likelihood-selected value of $q$ for the $b$th dataset.
Table~\ref{AIC} shows that the surrogate-likelihood-selected $\bar q$ closely tracks the value of
$q$ achieving the smallest MSE across different tail regimes: smaller $\bar q$
is selected under heavier tails (small $\nu$), while larger $\bar q$ is selected
under lighter tails or near-Gaussian errors.
\end{example}

\begin{table}[tbp]
\centering
\caption{\footnotesize Comparison of ARSPE--AIC and regression--scale polynomial estimators with fixed $q\in\{1,\ldots,10\}$ under Student's $t$ errors and $N(0,1)$ errors.}
\resizebox{0.9\linewidth}{!}
{
\begin{tabular}{lccccccccccc}
\toprule
\multicolumn{12}{c}{Student's $t$ with $\nu=1$}\\
\midrule
Performance & ARSPE--AIC & $q=1$ & $q=2$ & $q=3$ & $q=4$ & $q=5$ & $q=6$ & $q=7$ & $q=8$ & $q=9$ & $q=10$\\
\midrule
MSE  & \textbf{0.010} & \textbf{0.010} & 0.015 & 0.022 & 0.031 & 0.041 & 0.051 & 0.062 & 0.073 & 0.084 & 0.096 \\
Bias & \textbf{0.114} & \textbf{0.114} & 0.139 & 0.170 & 0.200 & 0.229 & 0.255 & 0.281 & 0.304 & 0.326 & 0.347 \\
$\bar q$ & 1.000 & -- & -- & -- & -- & -- & -- & -- & -- & -- & --\\
\midrule
\multicolumn{12}{c}{Student's $t$ with $\nu=1.5$}\\
\midrule
Performance & ARSPE--AIC & $\mathbf{q=1}$ & $q=2$ & $q=3$ & $q=4$ & $q=5$ & $q=6$ & $q=7$ & $q=8$ & $q=9$ & $q=10$\\
\midrule
MSE  & \textbf{0.009} & \textbf{0.009} & 0.011 & 0.013 & 0.016 & 0.019 & 0.021 & 0.024 & 0.026 & 0.028 & 0.030 \\
Bias & \textbf{0.109} & \textbf{0.109} & 0.116 & 0.130 & 0.143 & 0.155 & 0.165 & 0.174 & 0.182 & 0.189 & 0.195 \\
$\bar q$ & 1.100 & -- & -- & -- & -- & -- & -- & -- & -- & -- & --\\
\midrule
\multicolumn{12}{c}{Student's $t$ with $\nu=1.9$}\\
\midrule
Performance & ARSPE--AIC & $q=1$ & $q=2$ & $q=3$ & $q=4$ & $q=5$ & $q=6$ & $q=7$ & $q=8$ & $q=9$ & $q=10$\\
\midrule
MSE  & \textbf{0.009} & \textbf{0.009} & \textbf{0.009} & 0.011 & 0.012 & 0.014 & 0.015 & 0.016 & 0.017 & 0.018 & 0.019 \\
Bias & \textbf{0.108} & \textbf{0.108} & \textbf{0.108} & 0.117 & 0.125 & 0.133 & 0.138 & 0.143 & 0.148 & 0.152 & 0.155 \\
$\bar q$ & 1.775 & -- & -- & -- & -- & -- & -- & -- & -- & -- & --\\
\midrule
\multicolumn{12}{c}{Student's $t$ with $\nu=2.5$}\\
\midrule
Performance & ARSPE--AIC & $q=1$ & $q=2$ & $q=3$ & $q=4$ & $q=5$ & $q=6$ & $q=7$ & $q=8$ & $q=9$ & $q=10$\\
\midrule
MSE  & \textbf{0.008} & 0.009 & \textbf{0.008} & 0.009 & 0.010 & 0.010 & 0.011 & 0.011 & 0.012 & 0.012 & 0.013 \\
Bias & \textbf{0.099} & 0.105 & \textbf{0.099} & 0.104 & 0.109 & 0.114 & 0.117 & 0.120 & 0.123 & 0.125 & 0.126 \\
$\bar q$ & 2.007 & -- & -- & -- & -- & -- & -- & -- & -- & -- & --\\
\midrule
\multicolumn{12}{c}{Student's $t$ with $\nu=4.3$}\\
\midrule
Performance & ARSPE--AIC & $q=1$ & $q=2$ & $q=3$ & $q=4$ & $q=5$ & $q=6$ & $q=7$ & $q=8$ & $q=9$ & $q=10$\\
\midrule
MSE  & \textbf{0.007} & 0.009 & \textbf{0.007} & \textbf{0.007} & \textbf{0.007} & 0.008 & 0.008 & 0.008 & 0.008 & 0.008 & 0.008 \\
Bias & \textbf{0.094} & 0.105 & \textbf{0.094} & 0.095 & 0.096 & 0.098 & 0.099 & 0.101 & 0.101 & 0.102 & 0.103 \\
$\bar q$ & 3.240 & -- & -- & -- & -- & -- & -- & -- & -- & -- & --\\
\midrule
\multicolumn{12}{c}{Student's $t$ with $\nu=25$}\\
\midrule
Performance & ARSPE--AIC & $q=1$ & $q=2$ & $q=3$ & $q=4$ & $q=5$ & $q=6$ & $q=7$ & $q=8$ & $q=9$ & $q=10$\\
\midrule
MSE  & \textbf{0.005} & 0.008 & 0.006 & 0.006 & \textbf{0.005} & \textbf{0.005} & \textbf{0.005} & \textbf{0.005} & \textbf{0.005} & \textbf{0.005} & \textbf{0.005} \\
Bias & \textbf{0.083} & 0.104 & 0.087 & 0.084 & \textbf{0.083} & \textbf{0.083} & \textbf{0.083} & \textbf{0.083} & \textbf{0.083} & \textbf{0.083} & \textbf{0.083} \\
$\bar q$ & 9.986 & -- & -- & -- & -- & -- & -- & -- & -- & -- & --\\
\midrule
\multicolumn{12}{c}{Student's $t$ with $\nu=26$}\\
\midrule
Performance & ARSPE--AIC & $q=1$ & $q=2$ & $q=3$ & $q=4$ & $q=5$ & $q=6$ & $q=7$ & $q=8$ & $q=9$ & $q=10$\\
\midrule
MSE  & \textbf{0.005} & 0.008 & 0.006 & 0.006 & \textbf{0.005} & \textbf{0.005} & \textbf{0.005} & \textbf{0.005} & \textbf{0.005} & \textbf{0.005} & \textbf{0.005} \\
Bias & \textbf{0.083} & 0.104 & 0.087 & 0.084 & \textbf{0.083} & \textbf{0.083} & \textbf{0.083} & \textbf{0.083} & \textbf{0.083} & \textbf{0.083} & \textbf{0.083} \\
$\bar q$ & 9.993 & -- & -- & -- & -- & -- & -- & -- & -- & -- & --\\
\midrule
\multicolumn{12}{c}{$N(0,1)$}\\
\midrule
Performance & ARSPE--AIC & $q=1$ & $q=2$ & $q=3$ & $q=4$ & $q=5$ & $q=6$ & $q=7$ & $q=8$ & $q=9$ & $q=10$\\
\midrule
MSE  & \textbf{0.005} & 0.008 & \textbf{0.005} & \textbf{0.005} & \textbf{0.005} & \textbf{0.005} & \textbf{0.005} & \textbf{0.005} & \textbf{0.005} & \textbf{0.005} & \textbf{0.005} \\
Bias & \textbf{0.079} & 0.103 & 0.084 & 0.080 & 0.080 & \textbf{0.079} & \textbf{0.079} & \textbf{0.079} & \textbf{0.079} & \textbf{0.079} & \textbf{0.079} \\
$\bar q$ & 10.000 & -- & -- & -- & -- & -- & -- & -- & -- & -- & --\\
\bottomrule
\end{tabular}
}
\label{AIC}
\end{table}

\noindent 
\textbf{Finite-sample comparison across distributions.}
We evaluate the finite-sample performance of the proposed adaptive
regression--scale polynomial estimator (Adaptive RSPE) across different sample
sizes and error distributions. We generate $1000$ datasets from
$
y_i = \beta_0 + \beta_1 x_{i1} + \varepsilon_i,\ i=1,\ldots,n,
$
with $\beta_0=1$, $\beta_1=2$, $x_{i1}\sim N(0,1)$, and $n\in\{500,600,800\}$.
We consider five error settings:
\begin{enumerate*}[label=(\arabic*)]
\item \(\mathcal{N}(0,\sigma^2)\) with \(\sigma\in\{1,2,3\}\);
\item \(\mathrm{Laplace}(0,\sigma)\) with \(\sigma\in\{3.5,4,4.5\}\);
\item Student's \(t(\nu)\) with \(\nu\in\{1,1.2,1.4\}\);
\item \(\pi\,\mathrm{Laplace}(0,5)\allowbreak+ (1-\pi)\mathcal{N}(0,10^2)\) with
\(\pi\in\{0.7,0.8,0.9\}\);
\item \(\pi\,t(1.2)+(1-\pi)\mathcal{N}(0,2^2)\) with
\(\pi\in\{0.6,0.7,0.8\}\).
\end{enumerate*}

Case~1 examines efficiency under Gaussian errors, whereas Cases~2--5 assess
robustness under heavy-tailed and contaminated distributions.

We compare Adaptive RSPE with eight estimators: LTS~\citep{rousseeuw1984least},
LMS~\citep{rousseeuw1984least}, TFHuber~\citep{wang2021new}, Huber with
$\lambda=1.345$~\citep{huber1973robust}, Tukey with $\lambda=4.685$~\citep{tukey1960survey},
MM~\citep{yohai1987high}, S~\citep{rousseeuw1984robust}, and LSE. For Huber and
Tukey, the scale is estimated via MAD based on LTS. We set $q_{\max}=10$ and
use $\gamma_u=1.5$, $\hat{\beta}^{(0)}=(1,1)$, $\hat{\sigma}^{(0)}=0.5$,
$\phi_0=10^{-4}$, $\varepsilon'=10^{-3}$.

Under Gaussian errors, we report relative efficiency
$
\mathrm{RE}=\frac{\mathrm{MSE}(\tilde{\beta}_n)}{\mathrm{MSE}(T_n)},
$
where $\tilde{\beta}_n$ is LSE and $T_n$ is a competing estimator. We also report
absolute bias, the average selected $\bar q$, and the average lower bound of the
breakdown point,
$\bar{\xi}
=
\frac{1}{1000}
\sum_{b=1}^{1000}
\min\left\{
1-\frac{1}{2\widehat{q}_b}-\frac{p-1}{n},
\frac{1}{2\widehat{q}_b}
\right\}.$

For non-Gaussian errors, we report MSE, absolute bias, $\bar{\xi}$, and $\bar q$.
The results for the sample size \(n=500\) are reported in
Table~\ref{tab:robust_estimators1}, whereas the results for \(n=600\) and \(n=800\) are deferred to the Supplementary Material.

\begin{table}[tbp] 
\centering \caption{\footnotesize Comparison of the performance of robust estimators under Cases~1--5 with sample size $n=500$.}\renewcommand{\arraystretch}{0.7} \resizebox{0.990\linewidth}{!}{ \begin{tabular}{lcccccccccccccccc} \toprule \multicolumn{12}{c}{\textbf{Case 1: $N(0,\sigma^2)$}} \\ \midrule \multirow{2}{*}{Robust estimator} & \multicolumn{4}{c}{$\sigma=1$} & \multicolumn{4}{c}{$\sigma=2$} & \multicolumn{4}{c}{$\sigma=3$} \\ \cmidrule(lr){2-5} \cmidrule(lr){6-9} \cmidrule(lr){10-13} & $Eff$ & $Bias$ & $\bar{q}$ & $\bar{\xi}$ & $Eff$ & $Bias$ & $\bar{q}$ & $\bar{\xi}$ & $Eff$ & $Bias$ & $\bar{q}$ & $\bar{\xi}$ \\ \midrule Adaptive RSPE & \textbf{1.000} &\textbf{0.073} & 10 & 0.05 &\textbf{1.000} &\textbf{0.147} & 10 &0.05 &\textbf{1.000} &\textbf{0.220} & 10 & 0.05 \\ TFhuber~\citep{wang2021new} &\textbf{1.000} &\textbf{0.073} & -- & -- &\textbf{1.000} &\textbf{0.147} & -- & -- &\textbf{1.000} &\textbf{0.220} & -- & -- \\ LTS &0.703 &0.087 & -- & -- &0.703 &0.175 & -- & -- &0.703 &0.262 & -- & -- \\ LMS &0.101 & 0.229 & -- & -- &0.101 &0.458 & -- & -- &0.101 &0.687 & -- & -- \\ Huber estimator &0.961 &0.075 & -- & -- &0.961 &0.149 & -- & -- &0.961 &0.224 & -- & -- \\ MM &0.959 &0.075 & -- & -- &0.959 &0.149 & -- & -- &0.959 &0.224 & -- & -- \\ Tukey's biweight &0.958 &0.075 & -- & -- &0.958 &0.150 & -- & -- &0.958 &0.224 & -- & -- \\ S &0.295 &0.136 & -- & -- &0.295 &0.271 & -- & -- &0.295 &0.407 & -- & -- \\ LSE &\textbf{1.000} &\textbf{0.073} & -- & -- &\textbf{1.000} &\textbf{0.147} & -- & -- &\textbf{1.000} &\textbf{0.220} & -- & -- \\ \midrule 
\multicolumn{12}{c}{\textbf{Case 2: $\mathrm{Laplace}(0,\sigma)$}} \\ \midrule \multirow{2}{*}{Robust estimator} & \multicolumn{4}{c}{$\sigma=3.5$} & \multicolumn{4}{c}{$\sigma=4$} & \multicolumn{4}{c}{$\sigma=4.5$} \\ \cmidrule(lr){2-5} \cmidrule(lr){6-9} \cmidrule(lr){10-13} & $MSE$ & $Bias$ & $\bar{q}$ & $\bar{\xi}$ & $MSE$ & $Bias$ & $\bar{q}$ & $\bar{\xi}$ & $MSE$ & $Bias$ & $\bar{q}$ & $\bar{\xi}$ \\ \midrule Adaptive RSPE &\textbf{0.032} &\textbf{0.202} & 2.368 & 0.240&\textbf{0.042} &\textbf{0.231} & 2.368 & 0.240 &\textbf{0.053} &\textbf{0.260} & 2.368 & 0.240 \\ TFhuber~\citep{wang2021new} &0.048 &0.246 & -- & -- &0.062 &0.281 & -- & -- &0.079 &0.316 & -- & -- \\ LTS &0.035 &0.210 & -- & -- &0.046&0.239 & -- & -- &0.058 &0.269 & -- & -- \\ LMS &0.166 &0.461 & -- & -- &0.217 &0.527 & -- & -- &0.274 &0.593 & -- & -- \\ Huber estimator &0.034 &0.208 & -- & -- &0.044 &0.238 & -- & -- &0.056 &0.268 & -- & -- \\ MM &0.034 &0.210 & -- & -- &0.045 &0.239 & -- & -- &0.057 &0.269 & -- & -- \\ Tukey's biweight &0.034 &0.209 & -- & -- &0.045 &0.239 & -- & -- &0.057 &0.269 & -- & -- \\ S &0.040 &0.224 & -- & -- &0.052 &0.256 & -- & -- &0.065 &0.288 & -- & -- \\ LSE &0.048 &0.246 & -- & -- &0.062 &0.281 & -- & -- &0.079 &0.316 & -- & -- \\ \midrule 
\multicolumn{12}{c}{\textbf{Case 3: $t(\nu)$}} \\ 
\midrule \multirow{2}{*}{Robust estimator} & \multicolumn{4}{c}{$\nu=1$} & \multicolumn{4}{c}{$\nu=1.2$} & \multicolumn{4}{c}{$\nu=1.4$} \\ \cmidrule(lr){2-5} \cmidrule(lr){6-9} \cmidrule(lr){10-13} & $MSE$ & $Bias$ & $\bar{q}$ & $\bar{\xi}$ & $MSE$ & $Bias$ & $\bar{q}$ & $\bar{\xi}$ & $MSE$ & $Bias$ & $\bar{q}$ & $\bar{\xi}$ \\ \midrule Adaptive RSPE &\textbf{0.008} &\textbf{0.098} & 1.000 &0.498 &\textbf{0.008} &\textbf{0.101} & 1.000 & 0.498 &\textbf{0.007} &\textbf{0.095} & 1.014 &0.495 \\ TFhuber~\citep{wang2021new} &0.656 &0.853 & -- & -- &0.193 &0.485 & -- & -- &0.086 &0.321 & -- & -- \\ LTS &0.010 &0.111 & -- & -- &0.010 &0.112 & -- & -- &0.009 &0.105 & -- & -- \\ LMS &0.036 &0.214 & -- & -- &0.036 &0.212 & -- & -- &0.035 &0.208 & -- & -- \\ Huber estimator &0.013 &0.130 & -- & -- &0.012 &0.123 & -- & -- &0.010 &0.114 & -- & -- \\ MM &0.011 &0.117 & -- & -- &0.010 &0.114 & -- & -- &0.009 &0.107 & -- & -- \\ Tukey's biweight &0.011 &0.116 & -- & -- &0.010 &0.113 & -- & -- &0.009 &0.106 & -- & -- \\ S &0.009 &0.107 & -- & -- &0.010 &0.111 & -- & -- &0.009 &0.108 & -- & -- \\ LSE &11167.54 &12.398 & -- & -- &33.058 &1.932 & -- & -- &1.964 &0.755 & -- & -- \\ \midrule 
\multicolumn{12}{c}{\textbf{Case 4: $\pi\,\mathrm{Laplace}(0,5)+(1-\pi)N(0,10^2)$}} \\ \midrule \multirow{2}{*}{Robust estimator} & \multicolumn{4}{c}{$\pi=0.7$} & \multicolumn{4}{c}{$\pi=0.8$} & \multicolumn{4}{c}{$\pi=0.9$} \\ \cmidrule(lr){2-5} \cmidrule(lr){6-9} \cmidrule(lr){10-13} & $MSE$ & $Bias$ & $\bar{q}$ & $\bar{\xi}$ & $MSE$ & $Bias$ & $\bar{q}$ & $\bar{\xi}$ & $MSE$ & $Bias$ & $\bar{q}$ & $\bar{\xi}$ \\ \midrule Adaptive RSPE &\textbf{0.121} &\textbf{0.390} & 2.257 & 0.244 &\textbf{0.103} &\textbf{0.359} & 2.048 & 0.249 &\textbf{0.083} &\textbf{0.325} & 2.049 & 0.249 \\ TFhuber~\citep{wang2021new} &0.183 &0.483 & -- & -- &0.161 &0.449 & -- & -- &0.129 &0.405 & -- & -- \\ LTS &0.123 &0.394 & -- & -- &0.111 &0.374 & -- & -- &0.088 &0.333 & -- & -- \\ LMS &0.550 &0.838 & -- & -- &0.438 &0.752 & -- & -- &0.377 &0.700 & -- & -- \\ Huber estimator &0.129 &0.404 & -- & -- &0.111 &0.371 & -- & -- &0.089 &0.338 & -- & -- \\ MM &0.132 &0.408 & -- & -- &0.114 &0.377 & -- & -- &0.090 &0.340 & -- & -- \\ Tukey's biweight &0.130 &0.407 & -- & -- &0.114 &0.376 & -- & -- &0.090 &0.339 & -- & -- \\ S &0.131 &0.407 & -- & -- &0.114 &0.381 & -- & -- &0.094 &0.347 & -- & -- \\ LSE &0.183 &0.483 & -- & -- &0.161 &0.449 & -- & -- &0.129 &0.405 & -- & -- \\ \midrule  
\multicolumn{12}{c}{\textbf{Case 5: $\pi\,t(1.2)+(1-\pi)N(0,2^2)$}} \\ \midrule \multirow{2}{*}{Robust estimator} & \multicolumn{4}{c}{$\pi=0.6$} & \multicolumn{4}{c}{$\pi=0.7$} & \multicolumn{4}{c}{$\pi=0.8$} \\ \cmidrule(lr){2-5} \cmidrule(lr){6-9} \cmidrule(lr){10-13} & $MSE$ & $Bias$ & $\bar{q}$ & $\bar{\xi}$ & $MSE$ & $Bias$ & $\bar{q}$ & $\bar{\xi}$ & $MSE$ & $Bias$ & $\bar{q}$ & $\bar{\xi}$ \\ \midrule Adaptive RSPE &\textbf{0.012} &\textbf{0.125} & 1.531 & 0.366 &\textbf{0.011} &\textbf{0.118} &1.176 &0.454 &\textbf{0.010} &\textbf{0.111} &1.031 & 0.490\\ TFhuber~\citep{wang2021new} &0.089 &0.328 & -- & -- &0.110 &0.364 & -- & -- &0.145 &0.417 & -- & -- \\ LTS &0.014 &0.133 & -- & -- &0.013 &0.130 & -- & -- &0.012 &0.123 & -- & -- \\ LMS &0.062 &0.283 & -- & -- &0.053 &0.259 & -- & -- &0.047 &0.241 & -- & -- \\ Huber estimator &0.014 &0.133 & -- & -- &0.014 &0.132 & -- & -- &0.014 &0.132 & -- & -- \\ MM &0.013 &0.129 & -- & -- &0.013 &0.128 & -- & -- &0.012 &0.126 & -- & -- \\ Tukey's biweight &0.013 &0.128 & -- & -- &0.013 &0.128 & -- & -- &0.012 &0.125 & -- & -- \\ S &0.017 &0.146 & -- & -- &0.014 &0.134 & -- & -- &0.012 &0.123 & -- & -- \\ LSE &27.267 &1.272 & -- & -- &50.543 &1.553 & -- & -- &25.341 &1.615 & -- & -- \\ \bottomrule \end{tabular} } \label{tab:robust_estimators1} 
\end{table}

Across heavy-tailed and contaminated settings, Adaptive RSPE consistently
outperforms the competing methods. In particular, TFHuber can degrade markedly
under heavy tails because its data-driven robustification rule depends on $n$,
causing the method to drift toward LSE as $n$ increases. In contrast, Adaptive
RSPE adapts to tail heaviness through $\hat q$ without an explicit $n$-driven
degeneracy. Moreover, when the sample is in general position, the breakdown
lower bound $\bar\xi$ increases as the tails become heavier, reflecting an
intuitive robustness adaptivity.

\subsection{Applications} 
\subsubsection{Example A: Real Estate Valuation}

\label{subsec:Hous_price_uni_area}
We analyze the Real Estate Valuation dataset (\url{https://archive.ics.uci.edu/dataset/477/real+estate+valuation+data+set})
with response \textit{House price of unit area} and $6$ predictors. Figure~5(a) in the Supplementary Material
shows the histogram of House price of unit area, indicating small outliers in the response. We fit LSE, LMS, MM, Huber ($k=1.345$, MAD--LMS scale), Tukey ($k=4.685$,
MAD--LMS scale), S, and the proposed method. We use $q_{\max}=10$, $\gamma_u=1.5$,
$\hat{\beta}^{(0)}=(1,1,1,1,1,1)$, $\hat{\sigma}^{(0)}=0.5$, $\phi_0=10^{-4}$,
and $\varepsilon'=10^{-3}$. We evaluate prediction via bootstrap + train/test splitting: each replication
uses $300$ training observations and $114$ test observations. Table~\ref{tab:plasma_boston}
reports AMAPE and SE over $200$ replications, together with $\bar q$, $\bar{\xi}$ and $\bar\kappa$. The results indicate that the proposed estimator 
yields a substantially lower AMAPE than the seven competing robust methods.





\subsubsection{Example B: Building energy efficiency (UCI Energy Efficiency)}
\label{subsec:ex_energy_efficiency}

We analyze the Building energy efficiency dataset (\url{https://archive.ics.uci.edu/ml/datasets/energy%2Befficiency})
with response \textit{Heating load} and $4$ predictors, including relative compactness, surface area, wall area and roof area. Figure~5(b) in the Supplementary Material
shows the histogram of Heating load, indicating small outliers in the response. We fit LSE, LMS, MM, Huber ($k=1.345$, MAD--LMS scale), Tukey ($k=4.685$,
MAD--LMS scale), S, and the proposed method. We use $q_{\max}=10$, $\gamma_u=1.5$,
$\hat{\beta}^{(0)}=(1,1,1,1)$, $\hat{\sigma}^{(0)}=0.5$, $\phi_0=10^{-4}$,
and $\varepsilon'=10^{-3}$. We evaluate prediction via bootstrap + train/test splitting: each replication
uses $710$ training observations and $58$ test observations. Table~\ref{tab:plasma_boston}
reports AMAPE and SE over $200$ replications, together with $\bar q$, $\bar{\xi}$ and $\bar\kappa$. The results indicate that the proposed estimator 
yields a substantially lower AMAPE than the seven competing robust methods.

\subsubsection{Example C: Boston housing}

We analyze the Boston housing dataset (\url{http://lib.stat.cmu.edu/datasets/boston})
with response \textit{medv} and $13$ predictors. Following \citet{chang2018robust},
we focus on seven covariates---\textit{rm}, \textit{age}, \textit{dis}, \textit{tax},
\textit{ptratio}, \textit{black}, and \textit{lstat}. Figure~5(c) in the Supplementary Material
shows the histogram of housing prices, again indicating outliers in the response. We fit LSE, LMS, MM, Huber ($k=1.345$, MAD--LMS scale), Tukey ($k=4.685$,
MAD--LMS scale), S, and the proposed method. We use $q_{\max}=10$, $\gamma_u=1.5$,
$\hat{\beta}^{(0)}=(1,1,1,1,1,1,1)$, $\hat{\sigma}^{(0)}=0.5$, $\phi_0=10^{-4}$,
and $\varepsilon'=10^{-3}$. We evaluate prediction via bootstrap + train/test splitting: each replication
uses $400$ training observations and $106$ test observations. Table~\ref{tab:plasma_boston}
reports AMAPE and SE over $200$ replications, together with $\bar q$, $\bar{\xi}$ and $\bar\kappa$. The results indicate that the proposed estimator 
yields a substantially lower AMAPE than the seven competing robust methods.

\setcounter{figure}{1} 


\begin{table}[tbp]
\centering
\caption{Average median absolute prediction error (AMAPE) and standard
errors (SE) for the Real Estate Valuation, Energy Efficiency, and Boston
housing datasets.}
\label{tab:plasma_boston}

\begin{threeparttable}
\resizebox{\linewidth}{!}{
\begin{tabular}{lccccccc}
\toprule
\textbf{Dataset}
& \textbf{Proposed estimator} & \textbf{LMS}
& \textbf{Huber} & \textbf{MM} & \textbf{Tukey} & \textbf{S} & \textbf{LSE} \\
\midrule
Real Estate Valuation
& \textbf{4.489(0.558)}
&  4.910(0.707)
&  4.554(0.535)
&  4.494(0.519)
&  4.498(0.521)
&  4.861(0.596)
&  4.704(0.563)\\
Energy Efficiency
& \textbf{2.088(0.292)}
& 2.236(0.411)
&  2.108(0.311)
&  2.092(0.295)
& 2.089(0.300)
& 2.251(0.304)
& 2.165(0.363)\\
Boston
& \textbf{1.988(0.245)}
& 2.266(0.377)
& 2.043(0.226)
& 2.020(0.256)
& 2.026(0.252)
& 2.164(0.280)
& 2.273(0.289) \\
\bottomrule
\end{tabular}}

\begin{tablenotes}[flushleft]
\footnotesize
\item \parbox{\textwidth}{
Note: For the Real Estate valuation dataset, \(\bar q=2.485\),
\(\bar{\xi}=0.224\). For the energy efficiency
dataset, \(\bar q=3.24\), \(\bar{\xi}=0.219\).
For the Boston dataset, \(\bar q=1.945\), \(\bar{\xi}=0.263\).
}
\end{tablenotes}
\end{threeparttable}
\end{table}

\setcounter{figure}{1} 

\section{Conclusion}
We introduce a family of smooth $q$-polynomial losses for
joint regression--scale estimation. For every finite $q$, the loss
has logarithmic tails and a bounded, redescending regression score;
as $q$ grows, it approaches the squared loss pointwise. The
principal estimator uses the common calibration $\kappa=1/2$
and selects $q$ by maximizing a normalized surrogate likelihood.
This construction separates data-driven loss selection from
robustness-oriented tuning of the scale equation.

The scale score is bounded and its monotonicity leads to an exact
existence-and-uniqueness condition for the positive scale solution
conditional on the regression coefficient. The regression--scale
objective remains nonconvex in the regression parameters, so global
minimization and the behaviour of numerical stationary points must
be distinguished. For a finite candidate set and a uniquely
optimal surrogate-loss index, consistency of the profiled criteria
yields consistent selection and oracle equivalence for the
fixed-$\kappa$ estimator under the usual fixed-index regularity
conditions.

Fixing $\kappa=1/2$ does not guarantee a uniform high breakdown
point across $q$. In particular, the critical fraction for
fixed-regression scale explosion decreases as $1/(2q)$. A
breakdown-oriented alternative, $\kappa=q/2$, raises this
conditional threshold to $1/2$ but may alter the population scale
target and does not alone imply a $50\%$ joint breakdown point.
Proving sharp joint regression--scale breakdown bounds and studying
this optional adjustment, as well as computation from robust
initializations, are important directions for further work.

A remaining empirical step is to rerun the previous simulations
and applications using the fixed-$\kappa=1/2$ estimator, while
retaining the old results only as historical comparisons.

\section*{Supplementary Materials}

The supplementary material available upon request contains proofs of the theoretical results,
additional algorithmic details, and further simulation results.
\par
\par


\bibhang=1.7pc
\bibsep=2pt
\fontsize{9}{14pt plus.8pt minus .6pt}\selectfont
\renewcommand\bibname{\large \bf References}
\expandafter\ifx\csname
natexlab\endcsname\relax\def\natexlab#1{#1}\fi
\expandafter\ifx\csname url\endcsname\relax
  \def\url#1{\texttt{#1}}\fi
\expandafter\ifx\csname urlprefix\endcsname\relax\def\urlprefix{URL}\fi


\bibliographystyle{plainnat}
\bibliography{RefPapers}

\end{document}